\documentclass[journal]{IEEEtran}
\usepackage[utf8]{inputenc}

\usepackage{cite}
\usepackage[usenames,dvipsnames]{xcolor}
\usepackage{graphicx}
\usepackage{circuitikz}

\usepackage{nomencl}
\makenomenclature

\usepackage[hidelinks,bookmarks=false]{hyperref}
\usepackage{booktabs,colortbl}
\usepackage{multirow}
\usepackage[cmex10]{amsmath}
\usepackage{ mathrsfs }
\usepackage{bbold}
\usepackage{array}
\usepackage{multirow}
\usepackage{booktabs,colortbl}

\usepackage{siunitx}
\usepackage{amssymb}

\usepackage{tikz}
\usetikzlibrary{decorations.pathreplacing}
\usetikzlibrary{arrows,shapes,positioning}
\usetikzlibrary{patterns}
\usepackage{pgfplots}
\pgfplotsset{compat=newest}
\pgfplotsset{plot coordinates/math parser=false}
\newlength\figureheight
\newlength\matlabfigurewidth
\newlength\matlabfigurewidthA
\ifCLASSOPTIONcompsoc
 \usepackage[caption=false,font=normalsize,labelfont=sf,textfont=sf]{subfig}
\else
 \usepackage[caption=false,font=footnotesize]{subfig}
\fi

\newcommand{\argmin}[1]{\underset{#1}{\operatorname{arg}\,\operatorname{min}}\;}

\newcolumntype{L}[1]{>{\raggedright\let\newline\\\arraybackslash\hspace{0pt}}m{#1}}
\newcolumntype{C}[1]{>{\centering\let\newline\\\arraybackslash\hspace{0pt}}m{#1}}
\newcolumntype{R}[1]{>{\raggedleft\let\newline\\\arraybackslash\hspace{0pt}}m{#1}}

\begin{document}

\title{Unsupervised Disaggregation of PhotoVoltaic Production from Composite Power Flow Measurements of Heterogeneous Prosumers}

\author{Fabrizio~Sossan,~\IEEEmembership{Member,~IEEE,}
        Lorenzo~Nespoli,~\IEEEmembership{Member,~IEEE,}
        Vasco~Medici,~\IEEEmembership{Member,~IEEE,}
        and Mario~Paolone,~\IEEEmembership{Senior~Member,~IEEE,}% <-this % stops a space
\thanks{
F. Sossan and M. Paolone are with the Distributed Electrical Systems Laboratory at the Swiss Federal Institute of Technology of Lausanne (DESL, EPFL), L. Nespoli and V. Medici are with the ISAAC at the University of Applied Sciences and Arts of Italian Switzerland (SUPSI), CH. Emails: \{fabrizio.sossan,mario.paolone\}@epfl.ch, \{vasco.medici, lorenzo.nespoli\}.supsi.ch
}% <-this % stops a space
\thanks{This research received funding from the Swiss Competence Center for Energy Research (FURIES).}% <-this % stops a space
%\thanks{Manuscript received on xx 2017;}
}

\maketitle

\begin{abstract}
We consider the problem of estimating the unobserved amount of photovoltaic (PV) generation and demand in a power distribution network starting from measurements of the aggregated power flow at the point of common coupling (PCC) and local global horizontal irradiance (GHI). The estimation principle relies on modeling the PV generation as a function of the measured GHI, enabling the identification of PV production patterns in the aggregated power flow measurements. Four estimation algorithms are proposed: the first assumes that variability in the aggregated PV generation is given by variations of PV generation, the next two use a model of the demand to  improve estimation performance, and the fourth assumes that, in a certain frequency range, the aggregated power flow is dominated by PV generation dynamics.
These algorithms leverage irradiance transposition models to explore several azimuth/tilt configurations and explain PV generation patterns from multiple plants with non-uniform installation characteristics. Their estimation performance is compared and validated with measurements from a real-life setup including 4 houses with rooftop PV installations and battery systems for PV self-consumption.
\end{abstract}


\mbox{}
 
%\nomenclature{$P_k$}{Active power flow}
%\nomenclature{$h$}{Planck constant}
 
%\printnomenclature

\section*{Nomenclature and Acronyms}
\addcontentsline{toc}{section}{Nomenclature}
\begin{IEEEdescription}[\IEEEusemathlabelsep\IEEEsetlabelwidth{$~~~~~~$}]
\item[PV] Photovoltaic.
\item[GHI] Global Horizontal Irradiance.
\item[GNI] Global Normal Irradiance.
\item[nRMS] Normalized Root Mean Square Error.
\item[PCC] Point of Common Coupling.
\item[MPPT] Maximum Power Point Tracking.
\item[$k$] Discrete time index.
\item[$P_k$] Active power flow at the PCC at time interval $k$.
\item[$\mathcal{I}^{\text{\----}}_k$] GHI observation at time $k$.
\item[$j$] Index for panel tilt/azimuth configuration.
\item[$I^{\diagdown}_{jk}$] Estimated GNI corrected for temperature for configuration $j$ at time $k$.
\item[$\widehat{G}_k$] Estimated PV production at time interval $k$.
\item[$\widehat{L}_k$] Estimated demand at time interval $k$.
\item[$\alpha_j$] PV nominal capacity at configuration $j$.
\end{IEEEdescription}

\section{Introduction}
Incresed levels of distributed photovoltaic (PV) generation determine higher reserve requirements at the system level and violations of voltage and line ampacity constraints in distribution systems during peak production hours \cite{5625442, 7593314}.
Technical solutions envisaged to mitigate PV generation drawbacks are curtailment strategies, control of converters active/reactive power, PV self-consumption schemes and dispatch of local power flows according to network-safe power consumption trajectories (e.g. \cite{7014236, 6708466, vanhoudt2014actively, FASO_ISGT2013, luthander2015photovoltaic, 7569022, 7736082, 7572208, 7542590}). A requirement for the implementation of those strategies is the availability of real-time production measurements from PV facilities. Incidentally, these are also useful to train data-driven prediction models (e.g. \cite{7570246, 7372485}). However, such a precondition is not always met in real-life conditions because installations are not always monitored, and, even when they are, factors such as \emph{i)} privacy concerns, \emph{ii)} conflicts due to the different owners of the metering infrastructures, and \emph{iii)} lack of standards for monitoring and aggregation of measurements, and their communication, play against the possibility of collecting real-time PV production measurements. 

As an alternative to direct monitoring of PV systems, we consider in this paper the problem of disaggregating PV generation from the aggregated active power measurements of a group of prosumers. The estimation principle relies on modelling PV generation as a function of the global horizontal irradiance (GHI), assumed known from local measurements. Four estimation algorithms are proposed and compared: the first assumes that the variability in the aggregated power flow measurements are mostly given by variations of the PV generation, the second and third leverage a model of the demand to improve estimation performance, and the fourth assumes that there is a certain frequency range in which the aggregated power flow measurement is dominated by PV generation components. All four algorithms use a transposition model to project GHI into a number of pre-defined differently oriented tilted planes to explain production from sites with different configurations.
The algorithms are designed to be unsupervised, i.e., they do not require measurements of the PV power profiles to be trained.
The algorithms are tested with measurements from a real-life setup of four houses with monitored rooftop PV plants and grid-connected battery systems, enabling the testing of estimation performance even when the demand is correlated with PV generation.

Even if in the existing literature the problem of disaggregation has been extensively investigated for nonintrusive load monitoring (e.g., \cite{kolter2010energy, kim2011unsupervised}), its application to PV disaggregation was considered in \cite{DBLP:journals/corr/KaraRTACS16}, which develops estimators of the total PV generation using the active power profile of a nearby installation and GHI proxy measurements as explanatory signals. With respect to \cite{DBLP:journals/corr/KaraRTACS16}, we leverage PV transposition models to identify PV production patterns from installations with different tilt/azimuth configurations, a key factor in urban contexts where PV generation is from rooftop PV plants and tilt and azimuth configurations are dictated by roofs characteristics and might not be uniform.

The paper is organized as follows: Section \ref{sec:problem} states the problem, \ref{sec:algs} presents the disaggregation algorithms,  \ref{sec:casestudy} discusses procedures and measurements for validation, \ref{sec:results} presents and discusses performance, \ref{sec:conclusions} summarizes key results and contributions.

\section{Notation and Problem Statement}\label{sec:problem}
\subsection{Configuration of the system}
\begin{figure}[!h]
\begin{center}
\includegraphics[width=0.9\columnwidth]{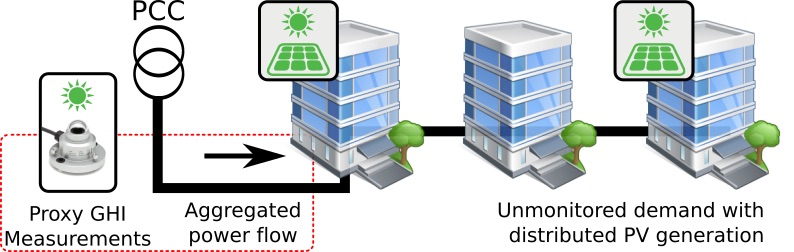}
\caption{A network topology with unmonitored demand and PV generation from multiple production sites with different azimuth/tilt configurations. The active power flow at the PCC and GHI are known from measurements. The problem is estimating the raw PV generation.}\label{fig:setup}
\end{center}
\end{figure}

We consider a feeder with distributed PV production, possibly from installations with different tilt and azimuth, and demand (e.g., Fig.~\ref{fig:setup}). The power injections at the single buses are not measured, however the total prosumption  (PV generation + demand) is known thanks to sensing the active power flow\footnote{Reactive power is not of special interest since PV plants normally operate at unitary power factor and, more in general, it is not possible to do assumptions on the kind of reactive power control implemented.} at the point of common coupling (PCC).  Local GHI values are from a pyranometer (although other methods could be considered, e.g. leveraging information available from nearby monitored PV installations \cite{scolari2017estimation}). We do the modeling assumption that the PV installations in the area are subject to the same GHI.
Local GHI measurements are known to be accurate in a range of 50 meters\cite{pelland2013photovoltaic}. Therefore, the proposed algorithms are expected to perform adequately when PV plants are spread over a small area, and their performance to decrease when considering larger areas.
Due to the small size of the networks that these methods target, grid transmission losses are neglected at this stage due to the short length of the cables.

\subsection{Notation}
The active power flow measured at the PCC at the discrete time interval $k$ is denoted as $P_k$ (\si{\kilo\watt}). Positive flows denote consumption and vice-versa (i.e., passive sign notation). GHI measurements are denoted by $\mathcal{I}^{\text{\----}}_k$ in \si{\kilo\watt\per\square\meter}, while $I^{\diagdown}_{jk}$ (\si{\kilo\watt\per\square\meter}) denotes the estimated global normal irradiance (GNI) to a certain tilted plane $j$ corrected for temperature (as described in \ref{sec:temperature}), where $j=1,\dots,J$ denotes the plane tilt/azimuth configuration.
We consider $J=21$ tilted planes with tilt and azimuth values equally spaced on a south-facing semi-sphere, chosen to have a reasonably representative set of potential configurations of PV installations in the northern hemisphere. GNI estimations are from the Hay-Davies transposition model \cite{hay1980calculation,maxwell1987quasi}. The quantity $\widehat{G}_k$ and $\widehat{L}_k$ (\si{\kilo\watt}) respectively denote  the estimated PV production and estimated demand, which are to be determined. A practical example of the disaggregation process is described in the following paragraph.

\subsection{Problem statement}\label{sec:problemstatement}
\begin{figure} [!h]
\centering
\includegraphics[width=1\columnwidth]{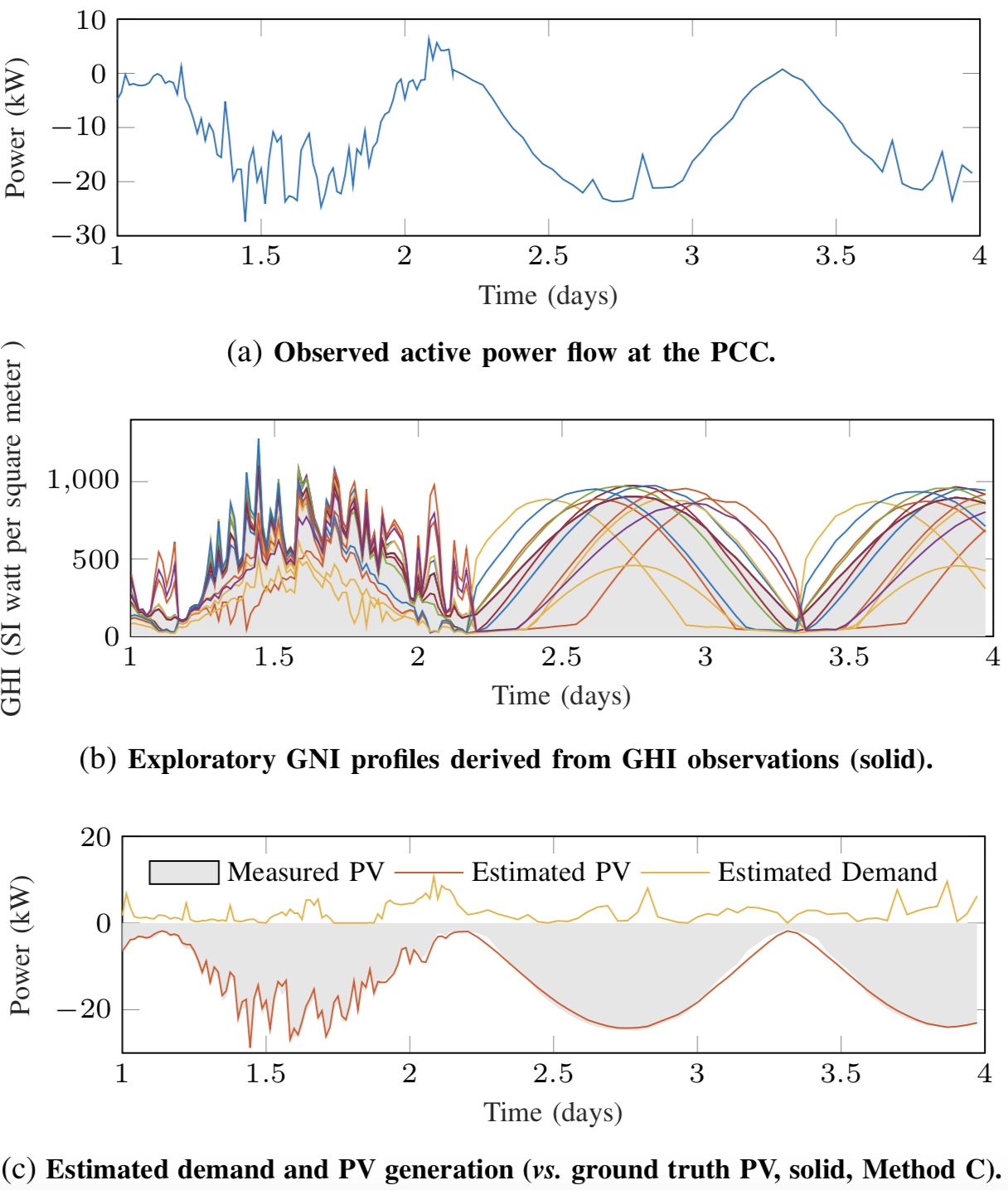}
%\subfloat[\scriptsize \bf Observed active power flow at the PCC.]
%{\scriptsize \input{fig/ex_Pagg.tex} \label{fig:ex_a}}~\\

%\vspace{-7mm}

%\subfloat[\scriptsize \bf Exploratory GNI profiles derived from GHI observations (solid).]
%{\scriptsize \input{fig/ex_irradiance.tex} \label{fig:ex_b}}~\\

%\subfloat[\scriptsize \bf Estimated demand and PV generation (\emph{vs.} ground truth PV, solid, Method~C).]
%{\scriptsize \input{fig/ex_estimated.tex}\label{fig:ex_c}}
\caption{Input, intermediate results and output of the proposed disaggregation  algorithms (night hours not shown).} \label{fig:ex}
\end{figure}

The problem is estimating the trajectories of the demand and total PV generation from measurements of the active power flow at the PCC and local GHI observations. This is exemplified in Fig.~\ref{fig:ex}a (night time observations are omitted) which shows the inputs, intermediate results and outputs of one among the proposed algorithms. The inputs are the aggregated power flow $P_k$ at the PCC (top panel) and GHI (middle panel, solid fill). The middle panel plot in Fig.~\ref{fig:ex}b also shows the GNI trajectories $I^{\diagdown}_{jk}$,  used to explore the potential PV production from plants with various tilt/azimuth configurations, as typical in urban feeders where panels are installed according to roof characteristics. Finally, the lower panel plot in Fig.~\ref{fig:ex}c shows the output of one of the proposed algorithms\footnote{Method~C, parameters $T_s=30$~s  $c=10$, mean nRMSE 5\%.}, with the estimated demand and estimated PV generation (orange line), the latter close to the measured ground truth PV generation (solid gray fill).

\section{Disaggregation Algorithms}\label{sec:algs}
The estimated global PV generation $\widehat{G}_k$ at the PCC is modelled as the sum over all the tilt/azimuth configurations $j=1,\cdots,J$ of the transposed irradiance $I^{\diagdown}_{jk}$ times $J$ nonnegative coefficients $\alpha_j \in \mathbf{R}_+$:
\begin{align}
 \widehat{G}_k(\boldsymbol{\alpha}) = \sum_{j=1}^{J}  \alpha_j \cdot I^{\diagdown}_{jk}, && k=1\dots,K, \label{eq:pvgeneration}
\end{align}
where $\boldsymbol\alpha = \{\alpha_j, j = 1,\cdots, J\} \in \mathbf{R}^J_+$ denotes the set of $\alpha_j$.
Physically, $\alpha_j$ is the PV generation capacity installed at configuration $j$.  It is measured in kWp, kilowatt peak, and denotes the amount of power produced in standard test conditions (STC, \SI{}{\kilo\watt\per\square\meter} GNI at \SI{20}{\celsius}). The temperature effect on the PV conversion efficiency is accounted for by preprocessing the input GNI time series as described in Section~\ref{sec:temperature}. By modelling the PV generation as in \eqref{eq:pvgeneration}, we assume that PV plants operate in the \emph{maximum power point tracking} (MPPT) mode; in case the output of a PV plant is controlled (i.e. curtailed), it is likely to be monitored and its contribution can be removed from the aggregated power flow, still allowing to apply the algorithms. Partial shading effects is not explicitly modeled, even if, as discussed in the following, some of the proposed methods are robust against it.

As detailed in the following, the proposed disaggregation algorithms estimate $\alpha$ (thus PV generation $\widehat{G}_k$) and require a training phase. Four algorithms, denoted as Method~A, B, C and D, are discussed. They attempt to estimate PV generation by exploiting different modeling principles inspired by the following empirical considerations:
\begin{itemize}
\item variations of PV generation dominate the variations of the power flow at the PPC. Method~A estimates PV generation by seeking for a trajectory with variations as close as possible to those observed at the PCC. The drawback is that variations are also due to demand changes (e.g. load inrushes);

\item the power flow at the PCC is modelled as the sum between PV generation and demand, where the latter is described by using a load model. Method~B and Method~C exploit two different load models, as explained later;

\item in a certain frequency range, the dynamics of the power flow at the PCC are dominated by those of PV generation. Method~D exploits the fact that demand and PV generation have different time dynamics: filtering the power flow measurements at the PCC makes possible to estimated the PV generation.
\end{itemize}
The validity of these empirical modelling considerations are tested in the results section by assessing and comparing the algorithms performance.

\subsection{Method A}
The unknowns $\boldsymbol{\alpha}$ are determined by assuming that the variability in the observed aggregated power flow are due to variations of the aggregated PV power. This is modeled by minimizing the norm-1 of the difference between the once differentiated time series $P_k$ and $\widehat{G}_k$ while subject to the estimated total PV production model \eqref{eq:pvgeneration}:
\begin{align}
\begin{aligned}
\boldsymbol{\alpha}^o = \argmin{ \boldsymbol{\alpha} \in \mathbf{R}^{J}_+}
\Bigg \{ \sum_{k=1}^{K} \bigg\vert & \left( P_{k} - P_{k-1} \right) + \\
-  &\left(\widehat{G}_k(\boldsymbol{\alpha}) - \widehat{G}_{k-1}(\boldsymbol{\alpha})  \right) \bigg\vert \Bigg\} \label{eq:methodB:2}
\end{aligned}
\end{align}
subject to:
\begin{align}
& \widehat{G}_k(\boldsymbol{\alpha}) = \sum_{j=1}^{J}  \alpha_j \cdot I^{\diagdown}_{jk}, && k=1\dots,K.\label{eq:methodB:3}
\end{align}
The problem in \eqref{eq:methodB:2}-\eqref{eq:methodB:3} is linear, thus convex and tractable.
The resolution of the input time series is a parameter of the Method~A, and its importance is discussed in the performance assessment. Method~A does not allow to model demand dynamics, e.g. load inrushes would be considered as variations of PV generation. The next two methods use a model of the demand to work around this limitation.

\subsection{Method~B}
Let $\widehat{\boldsymbol{L}} = [\widehat{L}_1, \dots, \widehat{L}_K]$ be the estimated demand trajectory. The estimated active power flow at the PCC $\widehat{P}_k$ is now written as:
\begin{align}
 \widehat{P}_k(\boldsymbol{\alpha}, \widehat{\boldsymbol{L}}) = \widehat{L}_k - \widehat{G}_k(\boldsymbol{\alpha}), && k=1\dots,K, \label{eq:estimatedaggregated}
\end{align}
i.e. the sum between the estimated total PV generation and demand, the former with negative sign because corresponding to generation. Method~B attempts to determine $\widehat{L}_k$ and $\boldsymbol{\alpha}$  by minimizing the norm-2 of the estimation error ${P}_k - \widehat{P}_k$. However, this problem is under-determined since the $K+J$ free variables are more than the number of observations $K$. Therefore, we augment the cost function and consider the sum of the least square and norm-1 of the once differentiated $\widehat{L}_k$ series (i.e., a combined linear regression and trend filtering problem, as for example in \cite{Kim2009} \cite{Li2015d}):
%The trend filtering problem can be also seen as a fused lasso regression with a zero penalty term \cite{Tibshirani2005}.
\begin{align}
\begin{aligned}
\begin{bmatrix}
 \boldsymbol{\alpha}^o \\
 \widehat{\boldsymbol{L}}^o
\end{bmatrix} = 
\argmin{
{\scriptsize
\begin{bmatrix}
\boldsymbol{\alpha} \in \mathbf{R}^{J}_+ \\
\widehat{\boldsymbol{L}} \in \mathbf{R}^{K}_+
\end{bmatrix}}
} \Bigg\{ & \sum_{k=1}^{K} \left ( P_k - \widehat{P}_k(\boldsymbol{\alpha}, \widehat{\boldsymbol{L}}) \right )^2 + \\
 + \lambda & \sum_{k=1}^{K} \left\vert \widehat{L}_k - \widehat{L}_{k-1} \right\vert  \Bigg\} \label{eq:methodB:0}
\end{aligned}
\end{align}
subject to:
\begin{align}
& \widehat{P}_k(\boldsymbol{\alpha}, \widehat{\boldsymbol{L}}) = \widehat{L}_k - \sum_{j=1}^{J}  \alpha_j \cdot I^{\diagdown}_{jk}, && k=1,\dots, K \label{eq:methodB:1}.
\end{align}
The cost function is the sum of a vector norm-1 and a quadratic cost function and convex if the latter term is convex. As shown in Appendix~A, the convexity of the quadratic term cannot depends on the input data and can be verified a-priori.

\subsection{Method C}
In Method~B, under-determination was solved by minimizing the norm-1 of the demand trajectory $\widehat{\boldsymbol{L}}$. As an alternative, we apply here a piecewise constant model of the demand, i.e. we require the unknown sequence $\widehat{\boldsymbol{L}}$ to be piecewise constant for $c$ consecutive samples, where $c$ is a parameter, by enforcing the following $c-1$ equality constraints:
\begin{align}
& \widehat{L}_{1} = \widehat{L}_{2} = \dots = \widehat{L}_{c},
\end{align}
for the case of the first $c$ samples. Extending to the set of $K$ measurements ($K$ multiple of $c$) yields:
\begin{align}
& \widehat{L}_{c(i-1) + 1}  = \dots = \widehat{L}_{c(i-1) + c}  && i=1,\dots,K/c.
\end{align}
Modelling the demand as piecewise constant is a reasonable assumption when the length of the constant segment does not overlap with typical intra-day dynamics of the demand, i.e. for small $c$ values and densely sampled series. In other words, it is reasonable when considering short periods of time (e.g., seconds), when the persistence model of the demand has unbeaten performance, see, e.g., \cite{hatziargyriou2014microgrids, Piga2016}. When the demand has shorter variations than $c$ (e.g., load inrushes), the estimated demand will have the average value of the waveform and the residuals will be the estimation error. The sampling time and $c$ are design parameters: the sensitivity of the algorithm performance with respect to their values is assessed in Section~\ref{sec:results}. Method~C consists in minimizing the norm-2 of the estimation error $P_k - \widehat{P}_k$ subject to the estimated aggregated power flow and piecewise constant demand models:
\begin{align}
\begin{bmatrix}
 \boldsymbol{\alpha}^o \\
 \widehat{\boldsymbol{L}}^o
\end{bmatrix} = 
\argmin{
\scriptsize \begin{bmatrix}
\boldsymbol{\alpha} \in \mathbf{R}^{J}_+ \\
\widehat{\boldsymbol{L}} \in \mathbf{R}^{K}_+
\end{bmatrix}
} \left\{ \sum_{k=1}^{K} \left ( P_k - \widehat{P}_k(\boldsymbol{\alpha}, \widehat{\boldsymbol{L}}) \right )^2  \right\} \label{eq:methodA:0}
\end{align}
%\beta \left| \left|
%\scriptsize\begin{bmatrix}
% \boldsymbol{\alpha} \\
% \widehat{\boldsymbol{L}}
%\end{bmatrix} \right|\right|_2
subject to:
\begin{align}
 & \widehat{P}_k(\boldsymbol{\alpha}, \widehat{\boldsymbol{L}}) = \widehat{L}_k - \sum_{j=1}^{J}  \alpha_j \cdot I^{\diagdown}_{jk}, && k=1,\dots, K \label{eq:methodA:1} \\
& \widehat{L}_{c(i-1) + 1}  = \dots = \widehat{L}_{c(i-1) + c}  && i=1,\dots,K/c \label{eq:methodA:2},
\end{align}
%where $||\boldsymbol{x}||_2$ in the cost function is a Lasso regularization term to preserve convexity, and $\beta$ is a scalar weight. The convexity of the problem in \eqref{eq:methodA:0}-\eqref{eq:methodA:2}, along with the justification on the criterion used to determine $\beta$, is discussed in Appendix~\ref{}. On the other hand, the sensitivity of the method performance with respect to the series sampling time and $c$ -- two important design parameters -- is assessed in Section~\ref{sec:results}.
 The additional constraints \eqref{eq:methodA:2} are linear and do not impact convexity, thus the same consideration as for Method~B applies.

\subsection{Method D}
Method~D splits the observed active power flow at the PCC by exploiting similarities between the signals $P_k$ and $\widehat{G}_k(\boldsymbol{\alpha})$ in a certain frequency range. This approach is motivated by having verified similarities in the spectral density of the measured aggregated power flow and measured PV generation (available from the test site) with the Welch's periodogram method for coherence analysis \cite{Welch1967}.
Method~D initially filters the input GNI $I^{\diagdown}_{jk}$ and aggregated power flow $P_k$ with a sixth order Butterworth band-pass filter, where the low and high cut-off frequencies are parameters that reflect the frequency range where the aggregated power profile and PV generation are similar. They are free parameters and the sensitivity of the algorithm performance to their values is investigated in Section~V.  Let $\mathcal{P}_k$ and $\mathcal{I}^{\diagdown}_{jk}$ respectively denote the above mentioned filtered version of the sequence $P_k$ and transposed irradiance $I^{\diagdown}_{jk}$. The vector of unknowns $\boldsymbol{\alpha}^o$ is computed by the following robust linear regression:
\begin{equation}
\boldsymbol{\alpha}^o = \argmin{
\boldsymbol{\alpha} \in \mathbf{R}^{J}_+
}
\left\{ \sum_{k=1}^{K}  \rho\left(\mathcal{P}_k - G_k\left(\boldsymbol{\alpha}\right)\right)  \right\}\label{eq:D0}
\end{equation}
subject to:
\begin{align}
& \widehat{G}_k(\boldsymbol{\alpha}) = \sum_{j=1}^{J}  \alpha_j \cdot \mathcal{I}^{\diagdown}_{jk}, && k=1\dots,K. \label{eq:D1}
\end{align}
where $\rho(\cdot)$ is the bisquare loss function, see \cite{Holland1977}, a nonconvex relantionship which gives less weight to extreme values in the cost function to be robust against outliers. The problem \eqref{eq:D0}-\eqref{eq:D1} is solved by applying an iterative least square approach with guaranteed covergence \cite{Huber2009}.

\subsection{Temperature correction}\label{sec:temperature}
To account for the dependency between PV conversion efficiency and temperature, GNI values are corrected to reflect temperature variations from the reference value $T_{\text{ref}}$ (\SI{25}{\celsius}) by using the empirical model proposed in \cite{kratochvil2004photovoltaic}:
\begin{align}
I^{\diagdown}_{jk} = \mathcal{I}^{\diagdown}_{jk}  \left[1 + \gamma(T_{\text{cell},k}-T_{\text{ref}}) \right]
\end{align}
where $\mathcal{I}^{\diagdown}_{jk}$ is the original irradiance value, and $T_{\text{cell},k}$ the cell temperature at time $k$, estimated as\cite{kratochvil2004photovoltaic}%(page 18)
\begin{align}
T_{cell,k} = T_\text{air} +\beta \mathcal{I}^{\diagdown}_{jk},
\end{align}	
where $T_\text{air}$ is the air temperature, assumed known from local measurements, and $\beta=\SI{3.78 e-2}{}$ and $\gamma=\SI{-4.3e-3}{}$ are empirical and plant specific values here assigned considering typical average values\footnote{$\beta$ is the average of values for the close roof mount and open rack configurations from \cite{kratochvil2004photovoltaic} %(page 19)
, and $\gamma$ the average of the values for polycrystalline modules from \cite{pvmodules}.}.

\section{Methods for performance evaluation} \label{sec:casestudy}
%\subsection{Metrics for performance evaluation}
The performance assessment of the disaggregation algorithms is performed by evaluating their ability to reconstruct the PV generation time series starting from the measurements of the power flow at the PCC and GHI. Let $e_k = G_k - \widehat{G}_k$ ($G_k$ and $\widehat{G}_k$ are the PV generation ground-truth value and estimation, respectively) be the estimation error. The performance metrics are: normalized root mean square of the estimation error (nRMSE) $\left(\frac{1}{G}\frac{1}{K}\sum_{k=1}^K e_k^2\right)^{1/2}$, normalized mean absolute error (nMAE) $\frac{1}{G}\frac{1}{K}\sum_{k=1}^K |e_k|$, and normalized mean error (nME) $\frac{1}{G}\frac{1}{K}\sum_{k=1}^K e_k$, where $G$ is the total installed PV capacity (35.3~kWp) and $K$ is the samples number in the testing data set.

\subsection{Data sets for training and testing}
Time series are GHI, power flow at the PCC and PV generation measurements for 1 year from the real-life test facility described in the next paragraph. The first 2 are used for the training, while the last as the ground truth value to test the estimation performance. To preserve daily dynamics of the signals, time series are divided into daily sequences, then randomly shuffled to avoid to train and test the algorithms on different periods of the year.
The series time resolutions are 10, 30, 60, 120, 300, 600 and 900~sec (downsampling by samples average) and are to assess algorithms performance with respect to sampling time. Selected resolutions include those normally implemented in existing metering systems, e.g., 900~s is the resolution of smart meters in Switzerland and Italy. Here, the intent is to verify whether  such a sampling time is enough for the proposed application, or if performance would benefit from more densely sampled data. 
Each of the 7 datasets at different resolution is further split into 3 sub-sequences to perform a three-fold cross-validation, i.e., for each resolution, the algorithms are trained on the first fold and tested on the remaining 2; the process is repeated for all the 7 datasets, each time testing the algorithms on the part of the data which is not used for the training. In total, each algorithm is trained and tested 3 times for each resolution, for a total of 21 training and test runs.
% It is to note that this process is not applied to Method~D, which uses the dataset sampled at largest frequency since it cannot improve its performance by downsampling the data.
Measurements refer to days with a uniform mix of sky conditions: partly cloudy, clear sky and overcast. Algorithms performance is tested both when there are batteries performing PV self-consumption and when not, thus allowing to account for the case when the demand is correlated with PV generation.

\subsection{Experimental Setup}
Measurements are from a real-life setup in the region of Basel, Switzerland, with 4 private households, each equipped with a rooftop PV installation with different characteristics, as reported in Table~\ref{tab:PVsites}. PV converters operate in MPPT mode at unitary power factor.  The total PV installed capacity is 35.3~kWp and the peak demand is 12~kWp.
The households are also equipped with grid-connected battery systems with bidirectional power converter to implement PV energy self-consumption policies (actuated at 5 minutes resolution). Batteries specifications are summarized in Table~\ref{tab:PVsites}. Battery injections are monitored. They are removed from the power flow at the PCC by algebraic difference, such that, in the following analysis, it is possible to consider two cases: with and without battery action (self consumption).

\begin{table}[!ht]
\vspace{-0mm}
% increase table row spacing, adjust to taste
\renewcommand{\arraystretch}{1.12}
\centering
\caption{PV and battery systems in the experimental setup}\label{tab:PVsites}
\begin{tabular}{| C{0.5cm} | C{0.8cm} | c | C{0.7cm} | C{1.0cm} | C{1.2cm}  |}
\hline
\scriptsize House ID & \scriptsize PV capacity (kWh) & Azimuth & Tilt (\SI{}{\degree}) & \scriptsize Distance from House 1 (m) & \scriptsize Battery rating (kVA/kWh)\\
\hline
1 & 10.0 & 95   & 14 & 0     & 3/8.8\\
2(a) & 7.2   & 187 & 36 & 100 & 3/4.4\\
2(b) & 3.5 & 266 & 40 & 100 & -- \\
3 & 8.0   & 187 & 40 & 260 & 3/8.8\\
4 & 6.6   & 180 & 24 & 170 & 3/4.4\\
\hline
\end{tabular}
\end{table}

\paragraph{PV and power flow measurements}
The power flow at the PCC is the sum of the four households flows, measured synchronously at 10~s resolution. Similarly, the global PV production (used as the ground truth value to validate the estimation performance of the algorithms) is the sum of the single PV facilities power injections, measured at the converter level.

\paragraph{Global horizontal irradiance measurements}
GHI measurements are from a pyranometer installed on the roof of the household ID1. The line distance of the remaining households from the GHI observation point is shown in the second to last column of Table~\ref{tab:PVsites}. All the measurements are synchronized and timestamped, and logged in a time series database.

\section{Results and Discussion}\label{sec:results}
For a visual exemplification of the disaggregation process, the reader is referred to Section~\ref{sec:problemstatement} and Fig.~\ref{fig:ex}.
In this section, we first assess the performance of the proposed methods individually. In \ref{sec:results:comparison}, we perform a joint performance assessment to compare the quality of the estimations of the different algorithms and support the assertion that they can be considered unsupervised. Key results are discussed and summarized in \ref{sec:disc}.  In \ref{sec:lowerpv}, the algorithms are tested in scenarios with a lower penetration of PV generation to verify if less prominent patterns of PV generation are detrimental to estimation performance. Finally in \ref{sec:computational}, we discuss the computational performance.

 \subsection{Method~A}
\begin{figure} [!h]
\centering
\includegraphics[width=0.85\columnwidth]{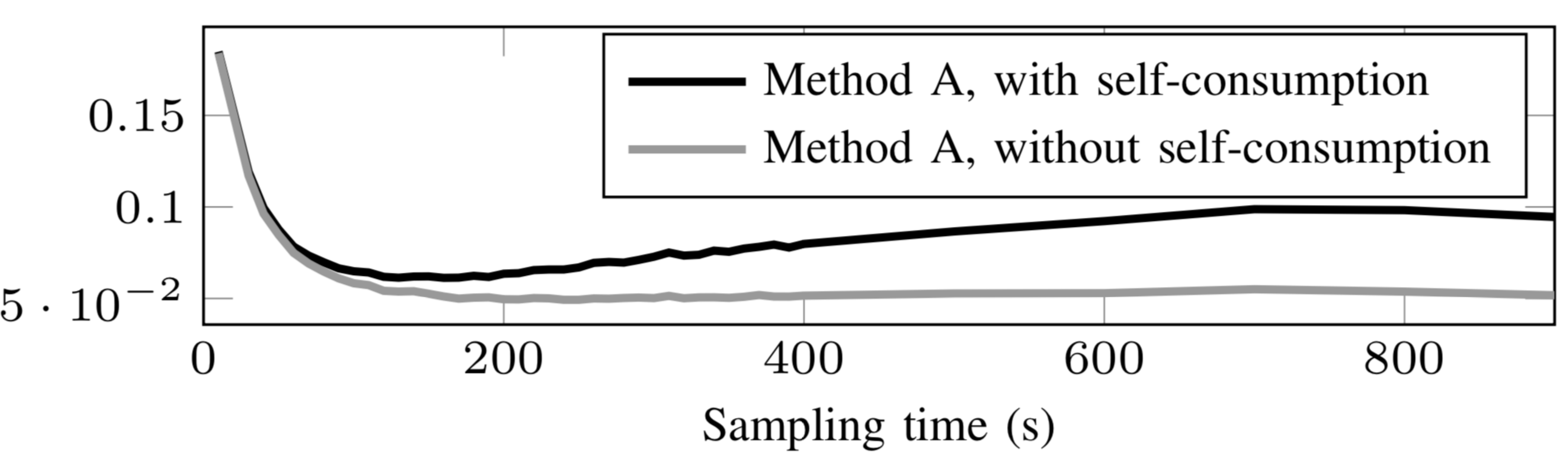}
\vspace{-2mm}
\caption{Method~A nRMSE as a function of the  input time series sampling time.}\label{fig:day0_ab}
\end{figure}

The estimation nRMSE as a function of the only parameter of Method~A (i.e. sampling time) with and without PV self-consumption is shown in Fig.~\ref{fig:day0_ab}. With no self-consumption, the nRMSE stabilizes at around 2~kW for sampling times larger than 200~s; with self-consumption, performance is poorer and is best at around 150~s. In both cases, estimation performance when the input time series is densely sampled (large sampling time) is poor.

\subsection{Method~B}\label{sec:results:B}
On top of the series sampling time, Method~B has a smoothing parameter $\lambda$ in \eqref{eq:methodB:0} to weight the demand time variations $\widehat{L}_k - \widehat{L}_{k-1}$ in the cost function. Performance as a function of the two parameters is shown in Fig.~\ref{fig:methodB}. With self-consumption, best performance happens in the middle right region of the parameter space. This region is larger in the case without self-consumption. Performance degradation patterns do not have a well identifiable trend.

\subsection{Method~C}\label{sec:results:C}
The parameters of Method~C are input time series sampling time and piecewise constant segment length $c$ (in number of samples). Their influence on the nRMSE is shown in Fig.~\ref{fig:methodC}. Estimation performance decreases when moving away from the axis origin, denoting that using densely sampled input time series and small $c$ values (the best performance is with 20~s resolution) are convenient. As mentioned in the formulation stage, this is to be expected because the choice of the two parameters affects the constant segment length of the demand piecewise constant model (i.e., the shorter it is, the better performance the persistence model has). %In particular, this assumption appears to be reasonable when considering short periods of time, i.e. where persistence model of the demand have unbeaten performance.
Estimation performance is worse with self-consumption (a numerical quantification is given in the next paragraph). Contour lines of Fig.~\ref{fig:methodC} denote that the performance degradation follows the same pattern when with and without self-consumption. Thus, even if estimation performance is different in the two cases, the optimal values of the parameters lay in the same parameters space area.

\subsection{Method~D}\label{sec:results:D}
\begin{figure*} [!h]
\centering
\subfloat[\scriptsize \bf nRMSE with self-consumption.] { \scriptsize
\includegraphics[width=0.85\columnwidth]{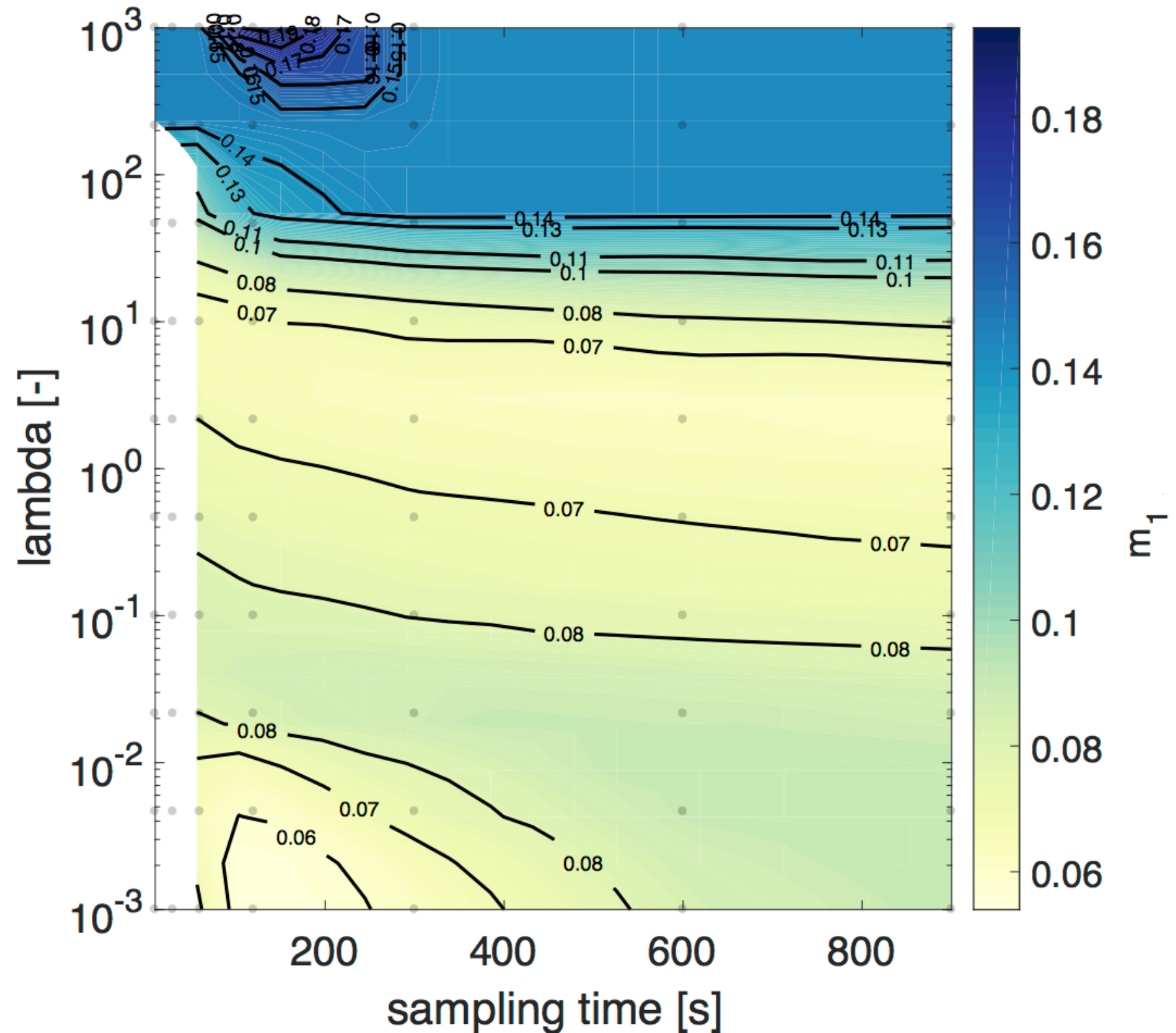}
\label{fig:methodB_self}}
\subfloat[\scriptsize \bf nRMSE without self-consumption.] { \scriptsize
\includegraphics[width=0.85\columnwidth]{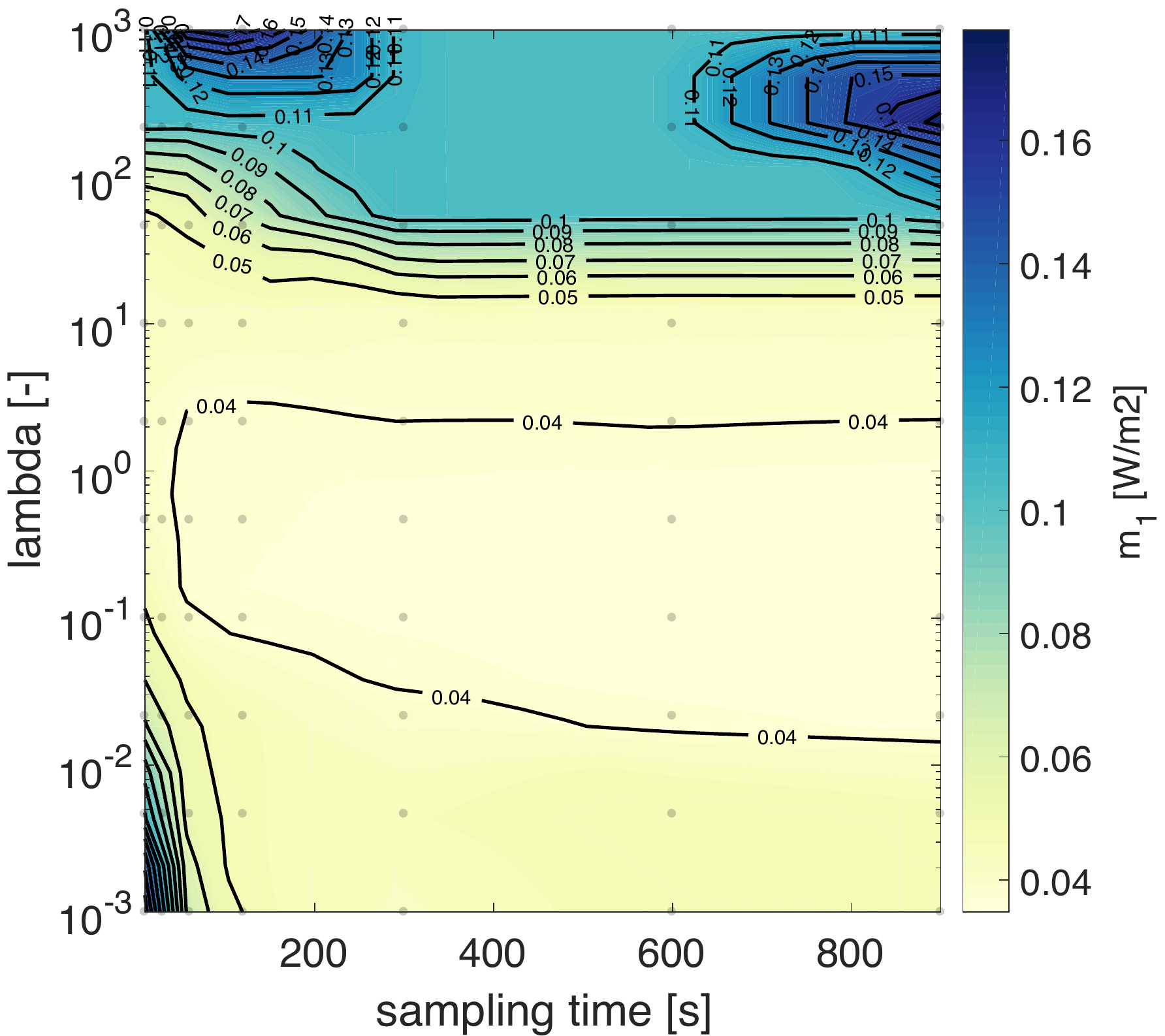}
\label{fig:methodB_noself}}
\caption{Method B sensitivity analysis: nRMSE as a function of the input time series sampling time and scaling parameter $\lambda$.} \label{fig:methodB}

\centering
\subfloat[\scriptsize \bf nRMSE with self-consumption.] { \scriptsize
\includegraphics[width=0.80\columnwidth]{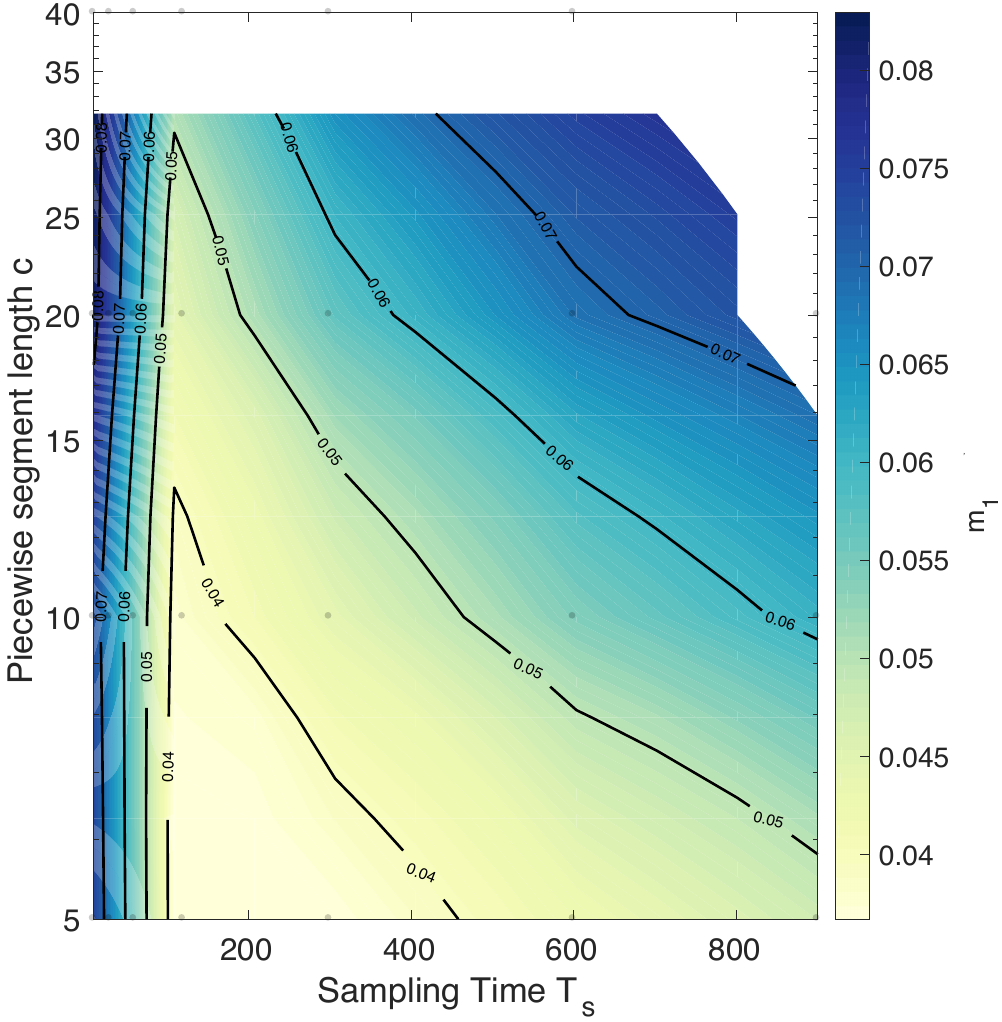}
\label{fig:methodC_self}}
\subfloat[\scriptsize \bf nRMSE without self-consumption.] { \scriptsize
\includegraphics[width=0.80\columnwidth]{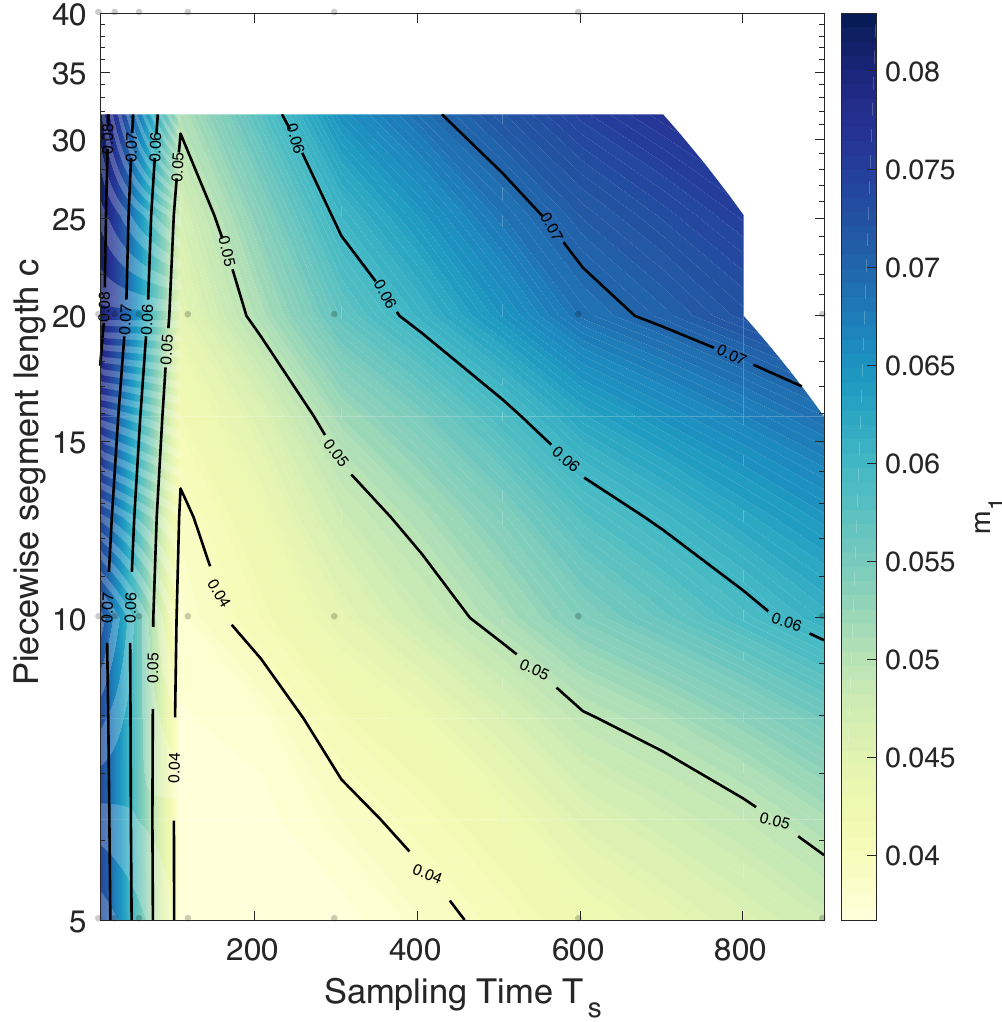}
\label{fig:methodC_noself}}
\caption{Method C sensitivity analysis: nRMSE as a function of the input time series sampling time and and length $c$ of the piecewise constant segment.} \label{fig:methodC}

\subfloat[\scriptsize \bf nRMSE with self-consumption.] { \scriptsize 
\includegraphics[width=0.85\columnwidth]{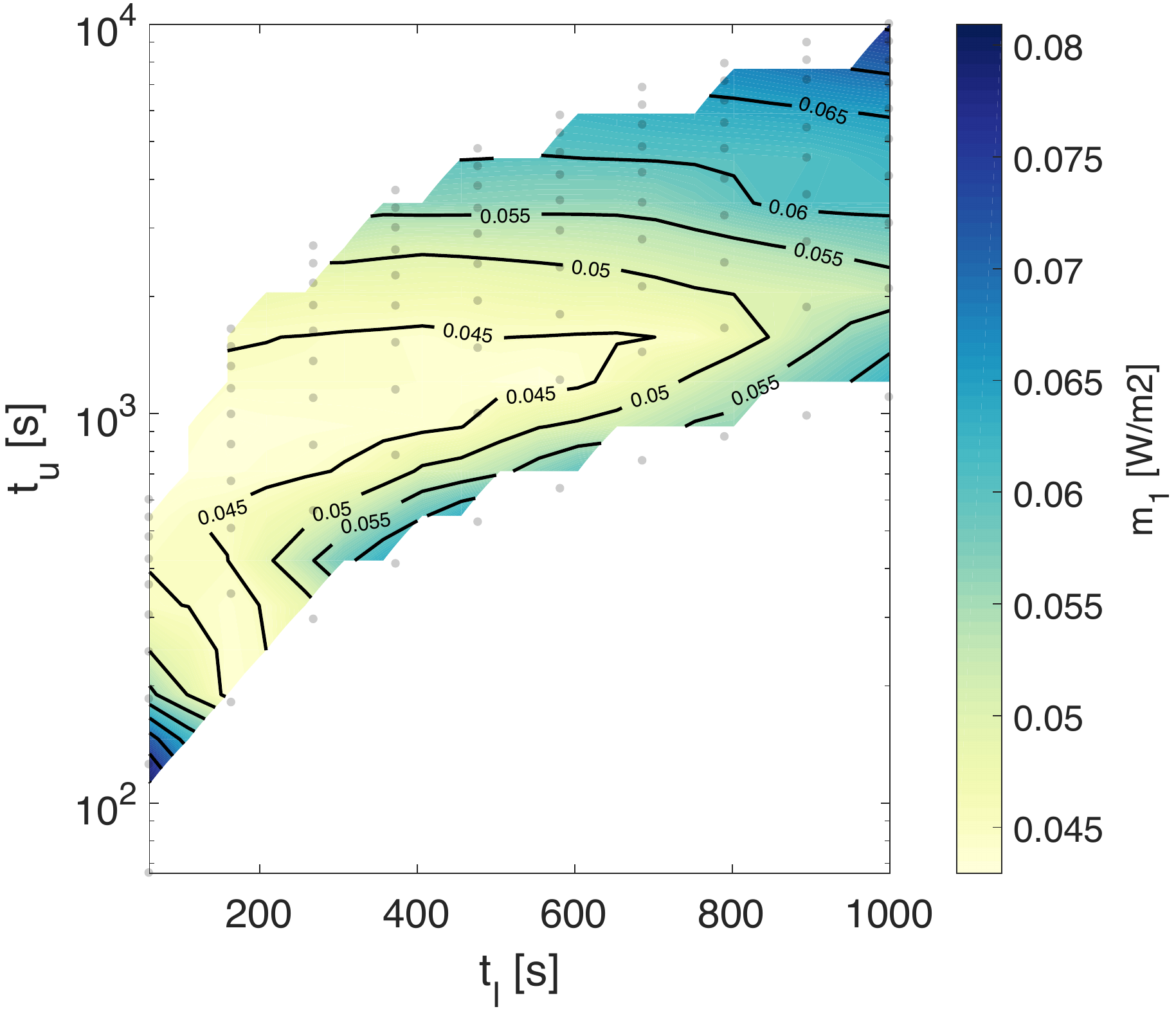}
\label{fig:methodD_self}}
\subfloat[\scriptsize \bf nRMSE without self-consumption.] { \scriptsize
\includegraphics[width=0.85\columnwidth]{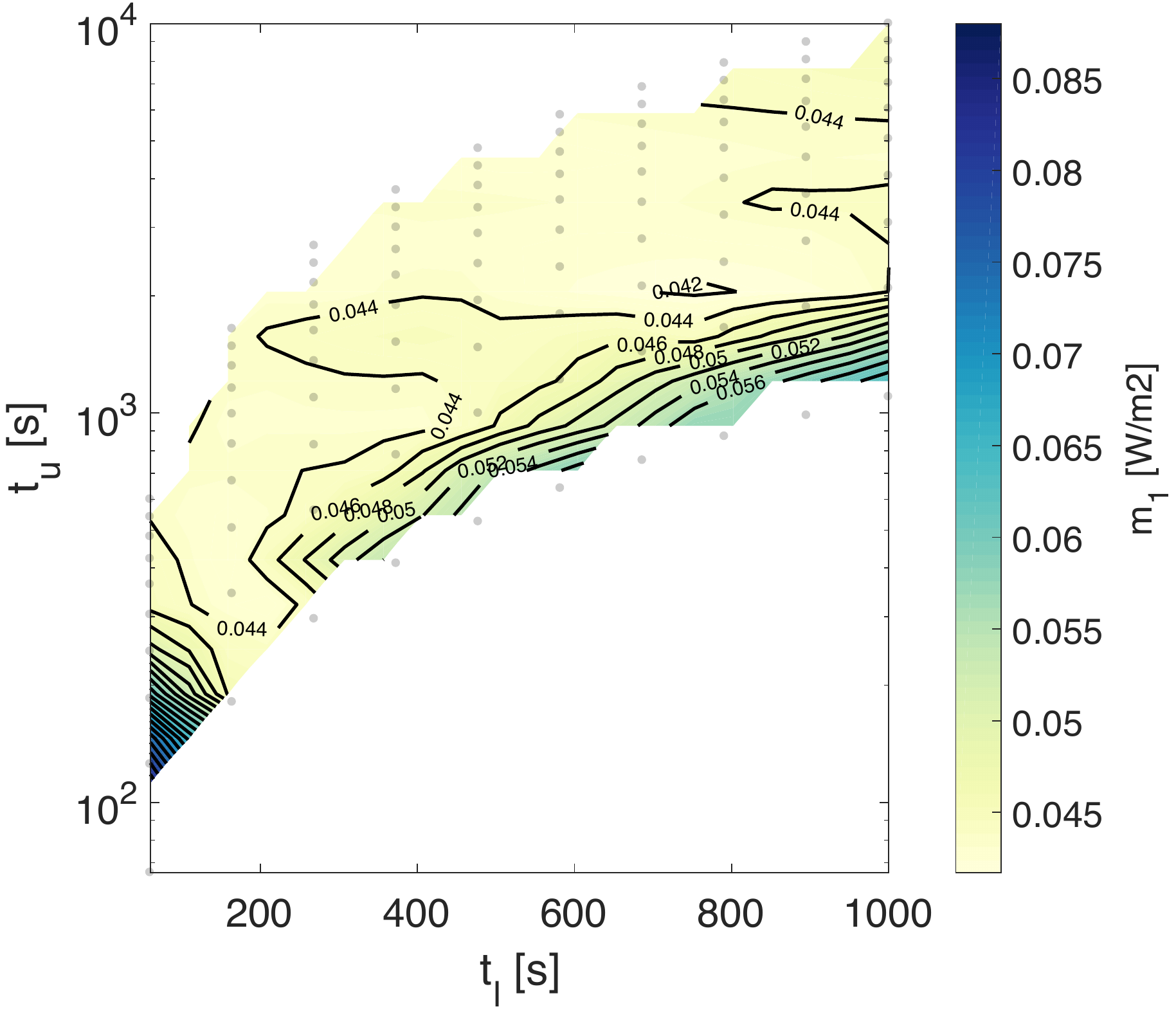}
\label{fig:methodD_noself}}
\caption{Method D sensitivity analysis: nRMSE as a function of the lower and upper cut-off frequencies. }
\label{fig:methodD}
\end{figure*}

Method~D parameters are the lower and upper cut-off frequencies of the bandpass filter, and a tuning constant of the bisquare loss function $\rho(\cdot)$. The last was found not to impact substantially on the algorithm performance and is therefore excluded from the current analysis. The sensitivity of algorithm performance to upper and lower cut-off frequencies is shown in Fig.~\ref{fig:methodD}. Best performance happens in a well identifiable region in the lower left part of the parameter space, which however tend to shrink in the case with self-consumption.

\subsection{Joint Performance Comparison}\label{sec:results:comparison}
The \emph{min, max, mean, and median} statistics of the estimation nRMSE, nMAE and nME of the 4 algorithms are reported in Table \ref{tab:RMSE}. For each algorithm, the reported statistics are calculated over all the combinations of the considered parameters values. They are to be interpreted in the following way:
\begin{itemize}
\item \emph{min}: performance to be expected assuming to know a-priori the best performing set of parameters.
\item \emph{max}: performance to be expected  when choosing the worst possible combination of parameters.
\item \emph{mean}: performance to be expected when choosing a random combination in the parameters space.
\item \emph{median}: to evaluate performance distribution skewness.
\end{itemize}
Table~\ref{tab:RMSE} shows that all the methods perform poorer under self-consumption regimes (e.g. mean nRMSE 4.6 to 5.6\% and 5.3 to 7\% for Method~D and Method~C, respectively), in other words when the demand includes a component anti-correlated with PV generation. In terms of \emph{mean} and \emph{median} statistics, Method~D scores the best metrics, followed by C, B and A. In terms of \emph{min} value, Method~C outperforms the other, except for for the cases nMAE and nRMSE without self-consumption, where Method~B is better, and nME with self-consumption, where Method~D is the absolute best for all the metrics.% Finally, in terms of the \emph{max} statistic Method~D and and C are usually the best performing, except for for the case nME without self-consumption, where Method~B scores the best.

\begin{table}[!ht]
\renewcommand{\arraystretch}{1.05}
\centering
\caption{Estimation performance statistics (\%)}\label{tab:RMSE}
\begin{tabular}{| C{1.1cm} | C{1.1cm} | C{1.1cm} | C{1.1cm} | C{1.1cm} |}
\hline
\bf Statistic & \bf A & \bf  B  & \bf C & \bf  D \\
\hline
\multicolumn{5}{|c|}{nRMSE with self-consumption} \\
\hline
min & 5.22 & 5.2 & \bf{4.0} & 4.3\\
max & 20.6 & 17.8 & 10.4 & \bf{8.2}\\
mean & 9.7 & 9.2 & 7.0 & \bf{5.5} \\
median & 8.4 & 8.4 & 7.0 & \bf{5.4} \\
\hline
\multicolumn{5}{|c|}{nRMSE without self-consumption} \\
\hline
min & 3.5 & 4.4 & \bf{3.4} & 4.1\\
max & 17.9 & 17.7 & 9.5 & \bf{8.8} \\
mean & 6.6 & 7.62 & 5.3 & \bf{4.6} \\
median & 4.6 & 4.8 & 4.7 & \bf{4.4} \\
\hline
\multicolumn{5}{|c|}{nMAE with self-consumption} \\
\hline
min & 3.3 & 3.3 & \bf{2.5} & 2.7 \\
max & 14.0 & 11.8 & 8.5 & \bf{5.1} \\
mean & 6.2 & 5.9 & 4.7 & \bf{3.4} \\
median & 5.3 & 5.3 & 4.4 & \bf{3.4} \\
\hline
\multicolumn{5}{|c|}{nMAE without self-consumption} \\
\hline
min & \bf{2.2}  & 2.8  & 2.3 & 2.5 \\
max & 12.1 & 11.8 & \bf{5.4} & 5.6 \\
mean & 4.3 & 4.9 & 3.2 & \bf{2.9} \\
median & 3.0 & 3.0 & 3.0 & \bf{2.7} \\
\hline
\multicolumn{5}{|c|}{nME with self-consumption} \\
\hline
min & -13.9 & -11.7 & -8.1 & \bf{-4.3} \\
max & \bf{-1.8} & -1.8   & 8 & 4.3 \\
mean & -5.7 & -5.2  & 2.3 & \bf{-1.0} \\
median & -4.7 & -4.8  & -2.8 & \bf{-1.3} \\
\hline
\multicolumn{5}{|c|}{nME without self-consumption} \\
\hline
min & -11.9 & \bf{-1.2} & -4.8 & -5.0\\
max & 1.9 & \bf{-1.1} & 4.6 & 4.0\\
mean & -2.2 & -3.8 & 0.8 & \bf{0.5}\\
median & -1.2  & -1.7  & -1.0 & \bf{0.2}\\
\hline
\end{tabular}
\end{table}

\subsection{Discussion}\label{sec:disc}
The previous results showed that the algorithms with the largest number of best scores is Method~D, followed by C, B and A. If only sparsely sampled power flow observations are available (such as those from smart meters, typically at 15 minutes resolution), Method~D should be selected because it keeps good performance even at low resolutions.
If densely sampled observations are available, the performance of Method~C and D are comparable. In this case, Method~C has two advantages: \emph{i}) parameters can be selected with an educated physical-based guess, \emph{ii}) degradation patterns are the same for both with and without self-consumption. %In other words, the choice of the parameters in case of Method~C is more intuitive. 

When selecting the two best performing Methods (C and D), the mean nRMSE is 7\% and 5.5\% and nME is rather small, 0.5 and 0.8\%. The latter metric is of importance because it denotes that estimators are almost unbiased, in other words, even if a single estimation in time is wrong, the estimated global PV production over a period is nearly correct.

It is worth noting that, in the proposed sensitivity analysis, PV measurements have been used to assess estimation performance. However, in practical applications when PV observations are not available, it is not possible to do so otherwise there would be no use for disaggregation algorithms. In case of Method~C (min/max nRMSE in the range $4.0\div 13.6$\%), parameters can be chosen with an educated guess in order to get closer to the best performance. As far as Method~D is concerned, the min/max nRMSE gap is smaller ($4.2\div 9.6$\%), and estimation performance is good over a wide area of the parameter space.
In other words, although it is not possible to derive analytical criteria for performing an a-priori assignment of the parameters, empirical results showed that, in the proposed case, parameters can be chosen in a wide range of values without a sensible deterioration of the performance.

\subsection{Extension to cases with lower PV generation levels}\label{sec:lowerpv}
In this section, the algorithms with the best performing parameters from the previous analysis are tested in scenarios with lower PV production capacity to verify how less prominent PV generation patterns are detrimental to their performance. In each scenario, synthetic time series of the power flow at the PCC are generated as the sum between the demand and a fraction of the original PV generation. Two additional scenarios are considered, with 50\% and 25\% of the original PV capacity (17.5 and 8.75~kW, respectively), which correspond to 146 and 76\% PV penetration levels (i.e. installed PV capacity over the observed peak demand), respectively. Results are reported in Table~\ref{tab:pvcur}. Results show that lower PV penetration levels affects estimation performance of methods B and C (their performance worsens approximately by a factor 2 with a quarter of PV generation), whereas Method~D is more robust and its performance is minimally affected.

\begin{table}[!ht]
\renewcommand{\arraystretch}{1.10}
\centering
\caption{Estimation nRMSE (\%) at different levels of PV penetration with self-consumption.}\label{tab:pvcur}
\begin{tabular}{| C{1.9cm} | C{1.4cm} || C{0.6cm} | C{0.6cm} | C{0.6cm} | C{0.6cm} |}
\hline
\bf Nominal PV Capacity (kWp) & \bf PV Penetration (\%) & \bf A & \bf  B  & \bf C & \bf  D \\
\hline
%\multicolumn{5}{|c|}{nRMSE with self-consumption} \\
%\hline
35.3 & 294 & 5.2 & 5.2 & \bf 4.0 & 4.3 \\
17.6 & 146 & 9.4 & 6.9 & 9.7 & \bf 4.6 \\
8.8 & 73 & 18.7 & 10.4 & 11.8 & \bf 5.2 \\
\hline
\end{tabular}
\end{table}

\subsection{Computational aspects}\label{sec:computational}
In the disaggregation and estimation process, the only required real-time operation is the computation of \eqref{eq:pvgeneration}, a cheap task which involves algebraic and trigonometric relationships.
Computing $\boldsymbol{\alpha}^o$ is a training process without real-time requirements which can be executed off-line with historical data.  The computational time is 183~s for Method~A, 709~s for B, 103~s for C and 67~s for D.\footnote{Computational times refer to a workstation equipped with an Intel Xeon processor running at 2.70~GHz with Matlab on a virtualized operating system. Method~C and Method~A,B and D were executed on two different machines, machine 1 and 2. The computation time of Method~C was adjusted by a factor $t_2/t_1$, where $t_1$ and $t_2$ is the computation time of a reference problem executed on machine 1 and 2, respectively.}

\section{Conclusions}\label{sec:conclusions}
The problem of disaggregating a sequence of active power flow measurements composed of unobserved PV generation and demand into the respective trajectories was considered.
Four disaggregation algorithms were discussed. They attempt to explain similarities between the time series of the aggregated and estimated PV generation, three in the time domain and one in the frequency domain. Estimation algorithms leverage GHI measurements transposed onto a number of tilted planes with the objective of explaining PV production patterns from sites with potentially different configurations (a key feature in urban/suburban context where PV generation is mostly from rooftop PV facilities with tilt/azimuth configurations dictated by roof characteristics). The effect of the air temperature was modelled by preprocessing GHI values with a model-based approach.
Algorithms require an offline optimization problem-based training phase with historical data. For three algorithms, the convexity of the underlying optimization problem, important to assure tractability, is verifiable a-priori by inspecting the input data.  Reconstructing the PV power output requires computing on-line an algebraic relationship and is suitable for implementation with deterministic deadlines and low processing power, in real-time target devices. Algorithm performance was tested with data from a real-life setup, with PV generation from multiple sites with different configurations, different demand profiles, and battery systems for PV self-consumption. Results show that the best performing algorithms estimate PV generation with a root mean square and mean estimation errors in the ranges $3.4\div 8.8$\% and $0.5\div 2.3$\%, respectively, and that performance is minimally affected by the level of PV penetration in the prosumption mix. The practical utility of the proposed algorithms is envisaged in the context of power and energy management of distributed energy resources and data-driven PV generation forecasting in those situations where information from PV plants is not available due to issues such as privacy concerns or lack of adequate communication infrastructures.

\appendices
%\section{Computation } \label{app:GNIs}

\section{On the convexity of Method~B and C} \label{app:convexity}
We discuss on the convexity of the problem \eqref{eq:methodA:0}-\eqref{eq:methodA:2}. Let $\boldsymbol{P}=[P_1, P_2, \cdots, P_K]$, 
$\boldsymbol{I}^\diagdown_j=[I^{\diagdown}_{j, 1}, \cdots, I^{\diagdown}_{jK}],\ j=1,\dots,J$, $M \in \mathbf{R}^{K \times J}= \begin{pmatrix} \boldsymbol{I}^{\diagdown}_1, \boldsymbol{I}^{\diagdown}_2, \ldots, & \boldsymbol{I}^{\diagdown}_J \end{pmatrix}$ the matrix obtained by stacking horizontally the GNI columns. The estimated total PV production  \eqref{eq:pvgeneration} is (matrix product) $\widehat{\boldsymbol{G}} = M \boldsymbol{\alpha}$, which replaced into \eqref{eq:estimatedaggregated} yields:
\begin{align}
 \widehat{\boldsymbol{P}} =  \widehat{\boldsymbol{L}} - M \boldsymbol{\alpha}
 =\begin{pmatrix} \mathbb{1}_{K\times K }, & -M \end{pmatrix}
 \begin{pmatrix}
  \widehat{\boldsymbol{L}} \\
  \boldsymbol{\alpha}
 \end{pmatrix} 
 = S \boldsymbol{x} \label{eq:aggregatedmodel_vector}
\end{align}
where $\mathbb{1}$ is the $K\times K$ identity matrix, $S=\begin{pmatrix} \mathbb{1}_{K\times K }, -M \end{pmatrix} \in \mathbf{R}^{K\times (K+J)}$ and $x=( \widehat{\boldsymbol{L}}, \boldsymbol{\alpha} )^T$. The least square cost \eqref{eq:methodA:0} is:
\begin{align}
J &= (\boldsymbol{P} - \widehat{\boldsymbol{P}})^T(\boldsymbol{P} - \widehat{\boldsymbol{P}}) = \boldsymbol{P}^T\boldsymbol{P} + \widehat{\boldsymbol{P}}^T\widehat{\boldsymbol{P}}
 - 2{\boldsymbol{P}}^T\widehat{\boldsymbol{P}} \label{eq:Ja},
\end{align}
Minimizing \eqref{eq:Ja} is the same as minimizing (minimization is invariant under sum with constants and scale factors):
\begin{align}
 J &= \widehat{\boldsymbol{P}}^T\widehat{\boldsymbol{P}} - 2{\boldsymbol{P}}^T\widehat{\boldsymbol{P}}= \left( S \boldsymbol{x} \right)^T S \boldsymbol{x} - 
 2 \boldsymbol{P}^T S \boldsymbol{x} = \\
 &=  \boldsymbol{x}^TS^TS\boldsymbol{x} - 2 \boldsymbol{P}^T S \boldsymbol{x}
 = \frac{1}{2}\boldsymbol{x}^TH\boldsymbol{x} - \boldsymbol{f}^T \boldsymbol{x}, \label{eq:quad_form0}
\end{align}
where \eqref{eq:aggregatedmodel_vector} is used, $H = S^TS$ and $\boldsymbol{f} = S^T \boldsymbol{P}$. Eq.~\ref{eq:quad_form0} is convex if $H$ is semidefinite positive. Since $H$ depends on input data, convexity cannot be enforced by construction, but it can be checked numerically. It was noted that adding a regularization term to the matrix quadratic matrix $H'=H + \beta\cdot\mathbb{1} $ ($\beta=\SI{1e-4}{}$) helps to achieve convexity while not impacting substantially on algorithms performance.

\bibliographystyle{IEEEtran}
\bibliography{biblio_fab}

%\begin{IEEEbiography}[{\includegraphics[width=1in,height=1.25in,clip,keepaspectratio]{fig/bio/sossan.jpg}}]
\begin{IEEEbiographynophoto}
{Fabrizio Sossan}
is an Italian citizen and was born in Genova in 1985. He got his M.Sc. in Computer Engineering from the University of Genova in 2010, and, in 2014, the Ph.D. in Electrical Engineering from DTU, Denmark. Since 2015, he is a postdoctoral fellow at the Distributed Electrical Systems Laboratory at EPFL, Switzerland. In 2017, he has been a visiting scholar at NREL, Colorado, US. His main research interest are modeling and optimization applied to power system.
%\end{IEEEbiography}
\end{IEEEbiographynophoto}

%\begin{IEEEbiography}[{\includegraphics[width=1in,height=1.25in,clip,keepaspectratio]{fig/bio/nespoli.jpeg}}]
\begin{IEEEbiographynophoto}
{Lorenzo Nespoli} received the M.Sc. degree in Energy Engineering from Politecnico di Milano in 2013. Since 2014 he works on thermodynamic simulations and electric grid optimization at SUPSI, as a scientific assistant. He is a Ph.D. candidate at the Ecole polytechnique federale de Lausanne at the IPESE department.
%\end{IEEEbiography}
\end{IEEEbiographynophoto}

%\begin{IEEEbiography}[{\includegraphics[width=1in,height=1.25in,clip,keepaspectratio]{fig/bio/medici.jpg}}]
\begin{IEEEbiographynophoto}
{Vasco Medici} received a M.Sc. in Micro-Engineering from the Ecole polytechnique federale de Lausanne and a Ph.D. from the Institute of Neuroinformatics at the University of Zurich and ETH. He actually leads the Intelligent Energy Systems Team at the Institute for Sustainability Applied to the Built Environment at SUPSI. He has experience in project management, system identification, algorithmics, modeling and simulation. He is the coordinator at SUPSI for the Swiss Competence Center for Energy Research on Future Swiss Electrical Infrastructure SCCER FURIES. His team runs a number of pilot projects in the field of smart grid, in close collaboration with industrial partners.
%\end{IEEEbiography}
\end{IEEEbiographynophoto}

%\begin{IEEEbiography}[{\includegraphics[width=1in,height=1.25in,clip,keepaspectratio]{fig/bio/paolone.jpg}}]
\begin{IEEEbiographynophoto}
{Mario Paolone} was born in Italy in 1973. He received the M.Sc. (Hons.) degree in electrical engineering and Ph.D. degree from the University of Bologna, Bologna, Italy, in 1998 and 2002, respectively. He was a Researcher of Electric Power Systems at the University of Bologna in 2005, where he was with the Power Systems Laboratory until 2011. In 2010, he was an Associate Professor with Politecnico di Milano, Milano, Italy. He is now a Full Professor with the Swiss Federal Institute of Technology, Lausanne, Switzerland, where he was the EOS Holding Chair of the Distributed Electrical Systems Laboratory. He has authored and co-authored more than 170 scientific papers published in reviewed journals and presented at international conferences. His current research interests include power systems with particular reference to real-time monitoring and operation, power system protections, power systems dynamics, and power system transients. He is the Secretary and a member of several IEEE and Cigré Working Groups. He was a recipient of the IEEE EMC Society Technical Achievement Award in 2013.
%\end{IEEEbiography}
\end{IEEEbiographynophoto}

\vfill

\end{document}